\documentclass[pre,showpacs,floats,twocolumn,floatfix,superscriptaddress]{revtex4-1}

\renewcommand{\vector}[1]{{\bf #1}}

\newcommand{\scalarProduct}[2]{#1\cdot#2}
\newcommand{\vectorValue}[1]{|#1|}

\newcommand{\coordinate}{x}

\newcommand{\speed}{u}

\newcommand{\velocity}{\vector{\speed}}

\newcommand{\speedOfSound}{c_{s}}

\newcommand{\ttime}{t}
\newcommand{\timestep}{\Delta \ttime}
\newcommand{\lattice}{\Delta \coordinate}

\newcommand{\meanFreePath}{\lambda}

\newcommand{\latticeDensity}{\rho}

\newcommand{\flowRate}{\dot{Q}}

\newcommand{\temperature}{T}

\newcommand{\pressure}{p}

\newcommand{\shearViscosity}{\nu}
\newcommand{\dynamicViscosity}{\eta}

\newcommand{\equilibriumIndicator}{{\rm eq}}

\newcommand{\distribution}{f}

\newcommand{\latticeSpeed}{c}
\newcommand{\latticeVector}{\vector{\latticeSpeed}}
\newcommand{\latticeIndex}{k}

\newcommand{\weight}{w}
\newcommand{\relaxationTime}{\tau}
\newcommand{\relaxationParameter}{\lambda}

\newcommand{\dimlessformat}[1]{{ #1}}

\newcommand{\mach}{\dimlessformat{Ma}}
\newcommand{\knudsen}{\dimlessformat{Kn}}

\newcommand{\latticeUnits}{\rm l.u.}

\newcommand{\suppressionCoefficient}{\mathcal{S}}
\newcommand{\concentration}{\chi}

\usepackage{longtable}
\usepackage{graphicx}

\usepackage[UKenglish]{babel}

\begin{document}


\title{Mesoscopic simulation of diffusive contaminant spreading in gas flows at low pressure}


\author{S. Schmieschek}
\affiliation{Centre for Computational Science, University College London, 20 Gordon Street, London WC1H 0AJ, United Kingdom}
\affiliation{Department of Applied Physics, Eindhoven University of Technology, P.O. Box 513, 5600MB Eindhoven, The Netherlands}

\author{D. K. N. Sinz}
\affiliation{Department of Applied Physics, Eindhoven University of Technology,  P.O. Box 513, 5600MB Eindhoven, The Netherlands}

\author{F. Keller}
\affiliation{Institute for Chemical Process Engineering, University of Stuttgart, B\"oblinger Str. 72, Geb. 78, 70199 Stuttgart, Germany}

\author{U. Nieken}
\affiliation{Institute for Chemical Process Engineering, University of Stuttgart, B\"oblinger Str. 72, Geb. 78, 70199 Stuttgart, Germany}

\author{J. Harting}
\affiliation{Helmholtz-Institute Erlangen-Nuremberg (IEK-11), Research Centre J\"ulich, F\"urther Strasse 248, 90429 Nuremberg, Germany}
\affiliation{Department of Applied Physics, Eindhoven University of Technology,  P.O. Box 513, 5600MB Eindhoven, The Netherlands}



\date{\today}

\begin{abstract}
Many modern production and measurement facilities incorporate multiphase
systems at low pressures. In this region of flows at small, non-zero Knudsen-
and low Mach numbers the classical mesoscopic Monte Carlo methods become
increasingly numerically costly. To increase the numerical efficiency of
simulations hybrid models are promising. 
In this contribution, we propose a novel efficient simulation approach for the
simulation of two phase flows with a large concentration imbalance in a low
pressure environment in the low intermediate Knudsen regime. Our hybrid model
comprises a lattice-Boltzmann method corrected for the lower intermediate Kn
regime proposed by Zhang et al. for the simulation of an ambient flow field. A
coupled event-driven Monte-Carlo-style Boltzmann solver is employed to
describe particles of a second species of low concentration. In order to evaluate the model, standard diffusivity and diffusion advection systems are considered.

\end{abstract}


\maketitle

\section{Introduction} 
\label{sec:contaminant:introduction}

With technological advances in the 20th century, such as miniaturisation of micro electrical and bio-medical devices~\cite{fissell_high_2011,karniadakis_microflows_2000}, vacuum technology~\cite{jousten_handbook_2008}, space exploration~\cite{bird_molecular_1994} and production techniques requiring low pressure environments~\cite{mertens_progress_2004,karniadakis_microflows_2000}, flow systems have become of interest in which the mean free path $\lambda$ of individual fluid particles is not negligible in comparison to the geometry length-scale $l$. Therefore, hydrodynamic models assuming a fluid continuum lose validity as non-equilibrium, kinetic effects gain importance. This transition is commonly quantified by the Knudsen number $\text{Kn} = \lambda/l$. In this publication, results of the development of a model for two phase flow systems with large concentration imbalances in complex geometries at low pressures are reported. These types of system are of importance e.g. for reduction of disturbances and quality improvement in the operation  of advanced optical systems used in microchip production~\cite{mertens_progress_2004,gallis_approach_2001} and chemical vapour deposition applications~\cite{burns_dynamical_1997}.

While numerous numerical approaches capable of simulating flows at finite Knudsen number, including molecular dynamics~\cite{liu_molecular_2010,li_molecular_2010}, direct numerical simulation including slip corrections up to second order~\cite{hadjiconstantinou_limits_2006}, discrete velocity methods~\cite{broadwell_study_1964}, lattice-Boltzmann methods~\cite{ansumali_entropic_2006,toschi_lattice_2005} and Direct Simulation Monte Carlo (DSMC)~\cite{bird_molecular_1994} exist, the latter is by far the most common and has been largely used to benchmark other approaches. Despite its success in modelling aerodynamics problems in the upper atmosphere it has however become evident that for the case of flows in the lower transient and slip regime the DSMC approach still requires a very large number of particles to properly resolve a given flow problem~\cite{oran_direct_1998}.

Being based on directly solving the Boltzmann formalism as well, lattice-Boltzmann
methods (LBM) too are in principle capable to describe gas flows in arbitrary
flow regimes~\cite{succi_lattice_2001}. The commonly used D3Q19 single relaxation time
lattice-Boltzmann or lattice BGK model dealing with a linear approach to a discretised Maxwell-Boltzmann equilibrium distribution is however originally set up to recover the Navier-Stokes equations~\cite{qian_lattice_1992}. To extend its
applicability to the simulation of low pressure gas flows in the slip flow regime of small non-negligible Knudsen numbers, boundary conditions can be extended to include slip at solid walls~\cite{succi_mesoscopic_2002,tang_lattice_2004,zhang_lattice_2005,toschi_lattice_2005}. For simulating flows up to moderate Knudsen
numbers, phenomenological higher order lattice-Boltzmann models introducing wall function formalisms are
used~\cite{shan_kinetic_2006,ansumali_kinetic_2002}. Due to the existing high density gradients and dominating thermal velocity effects
however, the LBM alone is not suited for the simulation of diffusive spreading of
low concentrated components in rarefied gases. On the other hand the assumed
low concentration of the second phase allows for a greatly simplified
simulation approach, fully resolving particles where necessary, while
maintaining computational efficiency~(detailed in Sec.~\ref{sec:method:sub:algorithm}).

In order to increase the numerical efficiency of simulations at low Mach
numbers/in the slip flow and transition regime, several approaches involving
the extension of existing models and (spatial) hybrid methods have been
published. Fan et al. developed a modified DSMC method, called Information
Preservation technique to reduce statistical fluctuations associated with DSMC
at low Mach numbers~\cite{fan_statistical_2001}. Chun and Koch reduced these
fluctuations by solving the linearized Boltzmann equation together with a
correction of the particle velocities after the collision
step~\cite{chun_direct_2005}. Hybrid approaches combining classical
Navier-Stokes solvers and DSMC are developed to reduce numerical
costs~\cite{roveda_hybrid_1998, wijesinghe_three-dimensional_2004}. In these approaches,
the computational domain is divided with respect to the local flow regimes and
the DSMC method is only used in regions where non-equilibrium effects are
dominant. Since the coupling between the grid-based Navier-Stokes solver and
the particle based DSMC method is problematic, Burt et al. replaced the
grid-based method with a simplified DSMC approach in the continuum
region~\cite{burt_hybrid_2009}. Besides these hybrid approaches with
increase in efficiency in mind other hybrid models have been developed to
achieve a higher precision in approximations to solutions of the Boltzmann
equation, such as e.g. MD-DSMC hybrid methods~\cite{nedea_hybrid_2005}. 

To extend the applicability of the hybrid approach to length and timescales immediately relevant to enginieering problems regarding low pressure flows used to reduce contaminant concentrations, in the following a new hybrid approach integrating LBM and DSMC
simulation elements in a shared simulation domain is detailed. Here, the lattice-Boltzmann
method as implemented in the application LB3D serves as an efficient (slip boundary augmented) Navier-Stokes solver
for flows in complex geometries whereas a Monte Carlo particle model is used
to study diffusion of the dilute contaminants throughout the system. The focus on contaminant transport in a low
pressure environment allows to simplify the model significantly as here the
global flow field can be considered without perturbation by the
contaminants. Thereby the computational cost as compared to the hybrid models given above is further reduced- the contaminant molecules are assumed to behave at equilibrium with the background-gas where the explicit simulation of particle representatives adds a thermodynamically consistent diffusion model to the athermal LB flow solver.

After further elaborating on aspects of this newly developed approach, contaminant transport aspects in terms of thermal velocity, diffusivity and advection diffusion problems are evaluated. Results of simulations performed in a model of an experimental setup are reported before we close with concluding remarks.

\section{Modelling aspects}
The model aims to simulate dilute contaminants in a homogeneous atmosphere. This situation arises when low pressurised gas flows are employed to flush systems containing certain contaminants. The dilution is very important for the simplification of the model to hold. In particular for such cases it is possible to neglect the action of the contaminant on the atmospheric gas.
The LBM with intermediate Knudsen regime boundary corrections is employed to simulate the background gas. Single local contaminants are simulated as particles, where the collision processes with the background gas are solved implicitly by performing a Monte Carlo algorithm where the equilibrium velocity is Galileian shifted to account for advection.

In effect, the simulation of the contaminant component is independent of the background gas model as long as the latter provides a (quasi-)static velocity field and thermodynamic state as input. In the remainder we briefly revisit the modifications made to the LBM and give an overview of the Monte Carlo algorithm. Finally, the parameterisation is discussed in the context of an example thermodynamic configuration.

\subsection{Lattice-Boltzmann methods for intermediate Knudsen flow}

The lattice-Boltzmann (LB) approach is based on the Boltzmann kinetic equation
\begin{equation}
\label{eq:boltzmann}
\left[\frac{\partial }{\partial t}+{\bf u} \cdot \nabla_{\bf r}\right] f({\bf r,u},t)={\bf \Omega},
\end{equation}
which describes the evolution of the single particle probability density
$f({\bf r},{\bf u},t)$, where ${\bf r}$ is the position, ${\bf u}$ the
velocity, and $t$ the time. The derivatives on the left hand side represent
propagation of particles in phase space whereas the collision operator ${\bf
\Omega}$ takes into account molecular collisions.

In the LB method the time $t$, the position ${\bf r}$, and the
velocity ${\bf u}$ are discretised. In units of the lattice constant $\Delta x$
and the time step $\Delta t$ this leads to a discrete version of
Eq.~(\ref{eq:boltzmann}):
\begin{equation}
\begin{array}{cc}
f_k({\bf r}+{\bf c}_k, t+1) - f_k({\bf r},t) =
\Omega_k, &  k= 0,1,\dots,B.
\end{array}
\end{equation}
Our simulations are performed on a three dimensional lattice with $B=18$
discrete velocities in direction of face and edge centres of a cube and a rest-vector of zero velocity per site (the so-called D3Q19 model). With a proper choice of the discretised collision operator ${\bf \Omega}$ it can be shown that the flow behaviour follows the Navier-Stokes equation~\cite{succi_lattice_2001}.  We choose the Bhatnagar-Gross-Krook (BGK) form~\cite{bhatnagar_model_1954}
\begin{equation}
\label{eq:collision}
 \Omega_k =
 -\frac{1}{\tau}\left(f_k({\bf r},t) - f_k^{Eq}({\bf u}({\bf r},t),\rho({\bf r},t))\right)\mbox{ ,}
\end{equation}
which assumes a relaxation on a linear time-scale towards a discretised local Maxwell-Boltzmann distribution $f_k^{Eq}$. Here, $\tau$ is the mean collision time that
determines the kinematic viscosity $\nu=\frac{2\tau-1}{6}$ of the fluid.

The equilibrium distribution function $f^{Eq}$ chosen within the scope of this work is element-wise given by 

\begin{eqnarray}
\label{eq:feq}
\distribution_{\latticeIndex}^{\equilibriumIndicator} = \weight_{\latticeIndex} \latticeDensity \Big[ & 1 + \frac{1}{\speedOfSound^{2}}\scalarProduct{\latticeVector_{\latticeIndex}}{\velocity} + \frac{1}{2\speedOfSound^{4}} \left(\scalarProduct{\latticeVector_{\latticeIndex}}{\velocity}\right)^{2} - \frac{1}{2 \speedOfSound^{2}} \vectorValue{\velocity}^{2} + \nonumber \\ 
~& \frac{1}{2\speedOfSound^{6}} \left(\scalarProduct{\latticeVector_{\latticeIndex}}{\velocity}\right)^{3} - \frac{1}{2\speedOfSound^{4}} \left(\scalarProduct{\latticeVector_{\latticeIndex}}{\velocity}\right)\vectorValue{\velocity}^{2} \Big],
\end{eqnarray}
herein $g_k$ is a weight-factor depending on the lattice geometry and $\mathbf{c}_k$ are the lattice unit vectors and $\speedOfSound$ is the lattice speed of sound.

The particle velocity distribution $f$ can be related to physical properties via its moments. Here follows e.g. the macroscopic velocity vector $\mathbf{u}$ from the first order moment by

\begin{equation}
\label{eq:velocity}
\mathbf{u}=\frac{1}{\rho} \sum^{B}_{k=0} \mathbf{c}_k f_k(\mathbf{x},t),
\end{equation}

normalised by the local density

\begin{equation}
\label{eq:density}
\rho = \sum^{B}_{k=0} f_k(\mathbf{x},t),
\end{equation}

the zeroth order moment.

As detailed in the introduction, for flows in systems either at very low pressure or of geometry on the nanometer scale the characteristic dimensionless number, the Knudsen number $\text{Kn}$ becomes none negligible.  The mean free path for gases under ambient pressure $(1.013\; bar)$ is in the order of a few tens of nm, for classical technical flow problems in geometries much larger than this effects associated with higher Knudsen numbers can therefore be neglected.

For $\text{Kn} \; > \; 0.01$ rarefication effects gain noticeable influence on the flow characteristics and slip at the walls has to be taken into account~\cite{tang_lattice_2005}. This becomes relevant for macroscopically sized geometries in the regime of pressures of only a few Pascal considered here. Thus, to correct the lattice BGK (LBGK) Navier-Stokes solver for the discontinuous velocity field in the Knudsen layer, a slip boundary condition as proposed by Zhang {\it et al.}~\cite{zhang_lattice_2005} has been implemented. Other suitable slip boundary conditions were proposed by amongst others Tao {\it et al.}~\cite{tang_lattice_2005}, Ansumali and Karlin~\cite{ansumali_entropic_2006} and an extension of existing boundary conditions by Toschi and Succi introducing updated collision statistics by so-called virtual wall collisions~\cite{toschi_lattice_2005}.

In the remainder of this work, a two-step extension of the boundary condition by a diffusive reflection regime as well as a wall function modifying the effective mean free path is used as proposed by Zhang {\it et al.}. Following their formulation, the Knudsen number in a LBGK model is given by

\begin{equation}
\label{eq:KnudsenNumber}
 \text{Kn} = \sqrt{\frac{8}{3 \pi}} \frac{\tau-0.5}{L},
\end{equation}

where $L$ is the number of lattice sites used to resolve the characteristic length. From this it is immediately clear that the mean free path in this model is depending on the relaxation time $\relaxationTime$ only. It is thereby linked to the fluid viscosity. A first employed boundary condition allows to control the reflective behaviour and thereby implicitly the slip by an accommodation parameter $a$ defined between $a=0$ relating to no-slip or \textit{bounce-back} and $a=2$ implementing full-slip or \textit{specular reflection}. In the simulations described in the scope of this work, an accommodation parameter of $a=1$ is chosen. This corresponds to diffuse deflection at a rough wall~\cite{zhang_lattice_2005}.

As with higher $\text{Kn}$ and/or increase in resolution the Knudsen layer occupies a range of several lattice sites depth, simple slip boundary conditions overestimate the velocity at the boundary. The reason for this is a significant reduction in the mean free path of particles in the vicinity of a surface, effectively lowering the local Knudsen number. Recently, several methods have been introduced to reflect this by the introduction of an effective mean free path $\lambda_e$. Integrations with the lattice-Boltzmann method have been introduced by Tao {\it et al.}, Hyodo {\it et al.} and Zhang {\it et al.}~\cite{tang_lattice_2005, niu_kinetic_2007, zhang_capturing_2006}.\\

Here the correction proposed by Zhang {\it et al.} is applied, formulating the effective mean free path

\begin{equation}
\label{eq:effective-mean-free-path}
\lambda_e = \frac{\lambda}{1+0.7 e^{-\Delta y/\lambda}},
\end{equation}

that contains a dependence on the distance from the nearest boundary node $\Delta y$. Using Eq.~(\ref{eq:KnudsenNumber}) and Eq.~(\ref{eq:effective-mean-free-path}) the correction can be shown to correspond to a change in local kinematic viscosity, entering the model via a (now local) relaxation time parameter~\cite{zhang_capturing_2006}
\begin{equation}
 \label{eq:local-relaxation-time}
\tau= \sqrt{\frac{3 \pi}{8}}\left[ \frac{\lambda}{1+0.7 e^{-\Delta y/\lambda}} \right] +0.5.
\end{equation}

\subsection{A Monte Carlo Model for Contaminants}\label{sec:method:sub:algorithm}
To describe a second component present in very low concentrations only, a method derived from the DSMC approach is employed. As a solver to the Boltzmann equation, this model relies as well on the principal independence of the free movement and collision processes of particles.

DSMC algorithms solve the Boltzmann equation in terms of free movement and collisions of particles representative of a thermodynamical state. Free movement is calculated straightforward at a given (thermal) velocity assigned to the particles. The distinction of the method lies in its approach to model particle collisions. After free flight for the duration of a given collision interval particles are binned and randomised collision partners in a bin volume are drawn. Each of these pairs is subsequently calculated to perform a momentum conserving collision assigning a new particle velocity. This coarse grained collision model increases simulation efficiency dramatically as integration of the particle movement to predict explicit collisions is not necessary. At the same time the method has been shown to produce thermodynamically accurate results, especially in the context of dilute systems with finite Knudsen numbers.

In the approach presented here, the observation of collision time intervals is replaced by collision time calculation from the mean free path travelled in the system. Taking into account only atmosphere-contaminant interactions, atmospheric collision partners are not drawn from representative particles in a volume but rather created as pseudo-particles parameterised by properties of the LB field in the vicinity of a contaminant particle.

An event driven algorithm focusing on collision events is implemented. Assuming local equilibrium, according to the mass $m_{\zeta}$ of contaminant species $\zeta$ and temperature $T$ the individual particle velocity is drawn from a Maxwell distribution. From the mean particle velocity $u_{\zeta}$, the time span $\delta t$ to the next collision event is calculated as $\delta t = \lambda_{\zeta}/u_{\zeta}$, implicitly assuring the mean free path to be
\begin{equation}\label{eqLambda}
\lambda_{\zeta} = \frac{1}{\pi\sigma_{\alpha\zeta}^{2}n_{\alpha}} \sqrt{\frac{m^{\star}_{\alpha\zeta}}{m_{\zeta}}}.
\end{equation}
In this derivation, properties of atmospheric species $\alpha$ were used. The particle density per unit volume $n_{\alpha}$ is calculated for an ideal gas. The symbols $m^{\star}_{\alpha\zeta}=(m_{\alpha}m_{\zeta})/(m_{\alpha}+m_{\zeta})$
designate the reduced mass and $\sigma_{\alpha\zeta}$ the effective collision radius, calculated as the arithmetic average of atomic radii approximated by Lennard-Jones potential parameters given by Karniadakis \textit{et al.}~\cite{karniadakis_microflows_2000}.

At collision time the LBM lattice is used as binning grid. All contaminant particles present at a given lattice site undergo a collision with an ad hoc created \emph{pseudo particle}, generated as a representative of the local atmospheric gas flow. In particular, the atmosphere gas particles are assigned a velocity $\mathbf{u}_{\alpha}$ according to a Maxwell distribution whose mean value $\mathbf{\bar{u}}$ is shifted to reflect the velocity $\mathbf{u}_{\text{LBM}}$ of the lattice-Boltzmann field. Per direction $i$ this reads
\begin{equation}\label{eq.mawellDist}
u_{\alpha,i}= \sqrt{\frac{m_{\alpha}}{2 \pi k_B T}} \exp\left\{ \frac{-m_{\alpha}\left(\bar{u}_{i}-u_{\text{LBM},i}\right)^2}{2 k_B T}\right\}.
\end{equation}
Subsequent binary collisions between hard spheres are implemented classically, assuming conserved velocity of the centre of mass and thus global momentum conservation, while updating the relative velocity of the collision partners~\cite{bird_molecular_1994}.
The centre of mass velocity ${\bf u}_{m}$ and the relative velocity ${\bf u}_{r}$ of a contaminant particle $\zeta$ and a pseudo particle $\alpha$ are given by
\begin{equation}
 \label{eq:center_of_mass_vel}
{\bf u}_{m} = ({\bf u}_{\zeta} m_{\zeta}+{\bf u}_{\alpha}m_{\alpha})(m_{\zeta}+m_{\alpha})^{-1},
\end{equation}
and
\begin{equation}
  \label{eq:relative_vel}
  {\bf u}_{r} = {\bf u}_{\zeta}-{\bf u}_{\alpha},
\end{equation}
respectively.
With ${\bf u}_{m}^{*}$ and ${\bf u}_{r}^{*}$ denoting the post collision centre of mass and relative velocity respectively, kinetic energy and momentum is conserved in the collision by imposing ${\bf u}_{m}^{*} = {\bf u}_{m}$ and $|{\bf u}_{r}^{*}| = |{\bf u}_{r}|$.
Since particles are assumed to be hard spheres all directions for ${\bf u}_{r}^{*}$ are equally likely as scattering of hard spheres is isotropic.
The post collision velocities of the particles can then be determined using a randomly chosen direction for the relative velocity considering $|{\bf u}_{r}^{*}| = |{\bf u}_{r}|$ and the post collision equivalents of Eq. (\ref{eq:center_of_mass_vel}) and (\ref{eq:relative_vel}).
 The updated contaminant velocity is employed for the calculation of the time interval $\delta t$ elapsed until the next collision event. The information of the pseudo particle is discarded.
This collision process serves both to couple the contaminant particles to the atmospheric flow and as a thermostat (see section \ref{sec:valMaxwell}).

In the scope of this work the respective lattice-Boltzmann velocity field is pre-simulated until an equilibrium state is reached. There exists however no principle limitation on the synchronisation of the relaxation processes. Rather the quality of the modelling of dynamics in the style of DSMC is improved by taking into account the local time-dependent equilibrium determined by the LBGK algorithm. 
Analytical approximations of transport coefficients in this regime suggest e.g. that the mutual diffusivity can under these assumptions be expressed as a Lorentz-approximation (mass-ratio) corrected self-diffusivity coefficient of the atmosphere gas (see section \ref{sec:valFick}).

Rarefication effects are prominent in the Knudsen layer only. Thus special attention has to be paid to the boundary condition in the particle perspective as well. Since the main focus of the algorithm lies on true reproduction of the subscribed mean free path by timing collision events, the implementation of an accurate boundary condition is however straightforward. For the sake of simplicity again the lattice-Boltzmann site decomposition is used to define the boundary geometry, introducing discretisation errors in systems with curved boundaries. For these cases the method may be improved by another choice of surface definition, e.g. using interpolation techniques~\cite{ginzburg_multireflection_2003,chun_interpolated_2007}. However, it has been shown that by resolving the Knudsen layer by as little as four lattice sites the error observed in the flow field can be limited to the order of 5 percent. For a resolution of 8 lattice sites the deviation reduces to some per cent.
As discussed above for a working model of flow in the intermediate Kn regime surface slip has to be taken into account as well as the reduction in phase space volume in the vicinity of a boundary. Both requirements are met by a boundary condition reducing a particle's travelled distance if encountering a wall. 
The algorithm evaluates, whether during free movement a particle is travelling into a different LB node and subsequently whether it is a boundary node. If the particle would enter a boundary node, its position is reset to the point of contact, effectively reducing its travelled path. The subsequent collision event is modelled as diffusive reflection inverting the velocity component normal to the boundary surface. The thermal velocity is drawn from the half set with an inverted velocity component normal to the wall. This treatment is implicitly reducing the mean free path proportional to the surface to volume ratio in the system as well as reducing the phase space volume.
The boundary condition is evaluated for a simple advection diffusion problem for which an analytical solution is available (see section \ref{sec:valTransport}).

\subsection{Parameterisation}\label{sec:contaminantTransport:parametrisation}

The validity of the LBM is chiefly limited by the underlying (thermo-) statistical principles and approximations in expansions. It is possible to define a LB specific Knudsen limit based on the idea that the local equilibrium approximation on a lattice site does not hold anymore. Another way of putting this is given with the low Mach number limit which requires the \emph{transported} momentum density on the lattice to be low.
This clarifies that all modifications made to capture the intermediate Knudsen regime can only be phenomenological. As the evaluation of channel flow in the low Knudsen regime illustrates, modified continuum models can be suited to simulate flows in this regime. Nonetheless, when working with the modification of choice, leaving the core algorithm unmodified, the original restrictions have of course still to be observed.

In order to be able to understand the units captured by the model, a conversion of the core units is instructive. The scaling units of length $\Delta x$, time $\Delta t$ and mass $\Delta m$ are typically calculated by use of the constant speed of sound on the lattice as well as the kinematic shear viscosity imposed by the collision scheme. When aiming to model real systems, the usual choice of unit mass might be dismissed to calibrate the system in order to preserve numerical stability in cases were realistic pressures otherwise lead to very high densities or in cases where the relaxation rates are getting very low.

In the cases employed here, starting out from length-scaling the lattice to match a physical system, length and time scales are fixed by comparing the speed of sound in the desired physical system with the one on the lattice. Keeping the lattice mass at unity, the overall mass scaling is then determined by comparison of the resulting dynamic shear viscosity with the one of the desired physical system. 

A typical approach to the parameterisation of a simulation system is made by deciding on the spatial resolution of a system, determining the lattice spacing $\lattice$. Together with the lattice discretisation and the thereby determined speed of sound $\speedOfSound$ this immediately fixes the time step $\timestep$ = $\lattice/\speedOfSound$ and the conversion units for the kinematic viscosity $\shearViscosity$. For the case of a single fluid adhering to the ideal gas law $p = \rho c^2_{\textrm s}$ for the relation of pressure and fluid density, the consideration of the thermodynamic state and modelled substance, i.e. temperature, mass and pressure fixes a mass scale $\Delta m$. Conversion of the dynamic viscosity $\dynamicViscosity$ then allows to determine the relaxation time or shear viscosity relaxation parameter $\relaxationTime$ or $\relaxationParameter_{\shearViscosity}$ via  $\shearViscosity = \speedOfSound \timestep \left(\frac{\relaxationTime}{\timestep}-\frac{1}{2}\right)$. Here, some tuning flexibility is given by the mass scaling, allowing to adjust the simulated mean fluid density against the relaxation time parameter. Table~\ref{tab:conversion} gives some example numbers for arbitrary length scaling and Hydrogen at room temperature and $4$ Pascal pressure obtained from the NIST chemistry webbook~\cite{national_institute_of_standards_and_technology_thermophysical_2011}.

\begin{table*}

\begin{tabular}{| l | l | l |}
\hline
Property&Formulation& Physical example\\
\hline
Length scale $\lattice$& $\lattice = x^{\rm phys} / x^{\rm LB}$& $1~{\rm m} / 1000~\latticeUnits = 1.0 \cdot 10^{-3}$~m\\
\hline
Time scale $\timestep$& $\timestep = \speedOfSound^{\rm LB} / \speedOfSound^{\rm phys} \cdot \lattice$ &  $\lattice /\sqrt{3} / 1280 \frac{\rm m}{\rm s}  = 4.5105 \cdot 10^{-7}$~s \\
\hline
Mass scale $\Delta m$&$\Delta m = (\lattice^{3} \cdot \latticeDensity^{\rm phys})/\latticeDensity^{\rm LB}$& $(\lattice^{3} \cdot 2.445\cdot 10^{-6}~\frac{\rm kg}{\rm m^{3}})/0.1~\latticeUnits$ = $2.445\cdot 10^{-14}$~kg\\
\hline
Kinem. viscosity $\shearViscosity$& $\shearViscosity^{\rm LB} = \shearViscosity^{\rm phys} / \frac{\lattice}{\timestep^{2}}$ & $3.581~\frac{\rm m}{\rm s^2} / \frac{\lattice}{\timestep^{2}} = 1.6152 $\\
\hline
\end{tabular}
\caption{Example numbers for arbitrary length scaling and Hydrogen at room temperature and $4$ Pascal pressure obtained from the NIST chemistry webbook\label{tab:conversion}}
\end{table*}

\section{Validation}

We start out with the re-evaluation of the implementation of the intermediate Kn corrected LBM boundary conditions as described by Zhang \emph{et al.}~\cite{zhang_lattice_2005,zhang_capturing_2006}. Comparison is made to results of the linearized Boltzmann equations in a narrow channel published by Ohwada~\emph{et al.}~\cite{ohwada_numerical_1989,ohwada_numerical_1989-1}. Non-dimensionalised values are compared.
In a second part the thermal properties of the particle model as well as the coupling algorithm are tested. Throughout evaluations of the hybrid model SI units are used as they have been employed explicitly in the implementation of the particle model. Here, we compare the simulation velocity distribution to the respective Maxwell distribution both for the pseudo- and contaminant-particles over a range of temperatures.
Third, the diffusive behaviour of the model is checked. By focusing on a simple quasi-1D diffusion problem we ensure the validity of our choice of both mean free path $\lambda$ and diffusivity $D$. 
Finally the lattice-Boltzmann velocity field, the coupling and the diffusivity model are integrated with our contaminant boundary condition. Our results are compared to a solution of the transport equation in one dimension.

\subsection{Intermediate Knudsen numbers - The LBM implementation}
\label{sec:valLBM}
\begin{figure}
\center
\includegraphics[width=0.5\textwidth]{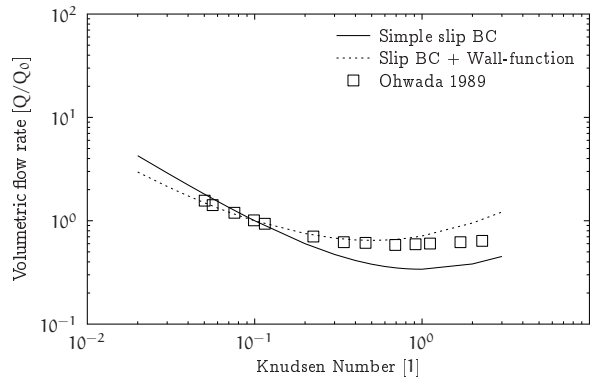}
\caption{Volumetric flow rate through a simple channel as a function of the Knudsen number, normalised by the flow rate measured at Kn=0.1. Reproduction of a validation published by Zhang \emph{et al.}~\cite{zhang_capturing_2006}. The simulation results are obtained in a channel of 32 lattice units width. A finite slip boundary condition was employed both alone and with a viscosity correction accounting for a varying mean free path in the vicinity of a boundary~\cite{zhang_lattice_2005,zhang_capturing_2006}. The reference values (symbols) are results of an exact solution to the linearized Boltzmann equation for hard spheres by Ohwada et al.~\cite{ohwada_numerical_1989,ohwada_numerical_1989-1}. Using the combined boundary conditions good agreement can be obtained for the lower intermediate regime of $Kn~\approx~0.05 .. 0.5$.
\label{fig:flowrates}}
\end{figure}

In order to assure the accuracy of the implementation of the boundary corrections by Zhang \emph{et al.}, flow at prescribed Knudsen numbers in a simple channel is simulated. Evaluations are made both for slip boundary conditions only, as introduced in~\cite{zhang_lattice_2005}, and for combinations of a slip boundary condition and a viscosity correction to account for a locally varying Knudsen number at the boundary~\cite{zhang_capturing_2006,sinz_simulation_2008}. Simulations are executed in a pseudo-1d channel of a size of 1x1x32 lattice units allowing for good numerical efficiency. The flow is driven by a body force of $F=1\cdot 10^{-7}$ in lattice units. To assure to reach a steady state simulations are in all cases run for 100,000 time steps. 

The flow profiles of the original publication are reproduced, reaching satisfactory agreement of simulated flow profiles and analytical solution to the linearized Boltzmann equation for hard spheres by Ohwada et al.~\cite{ohwada_numerical_1989,ohwada_numerical_1989-1}. As depicted in figure~\ref{fig:flowrates} the combined boundary corrections allow to recover the theoretical flow rates within a few percent even up to low single digit Kn. The shown flow rate has been normalised by the respective flow rate assumed at Kn=$0.1$. Furthermore the Knudsen-Paradox is captured by the model. The minimum flow rate measured in the simulations is reached around Kn$\approx 0.5$ in good agreement with the results of the analytical calculations as well as other numerical results reported by Toschi~\cite{toschi_lattice_2005} and Cercignani~\cite{cercignani_variational_2004}. These measurements allow for confidence in the capability of the extended lattice-Boltzmann model to capture fluid behaviour in the intermediate regime.

\subsection{The Maxwell speed distribution - The coupling algorithm}
\label{sec:valMaxwell}
\begin{figure}

\includegraphics[width=0.5\textwidth]{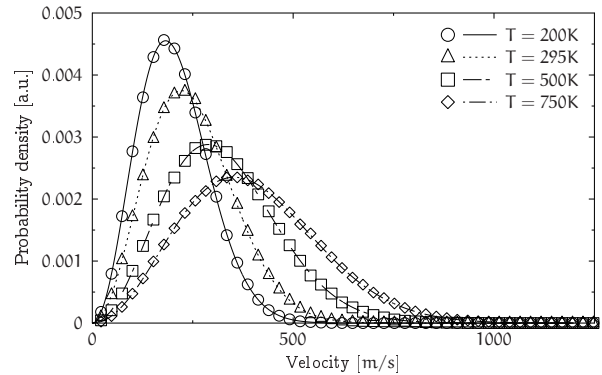}
\caption{Probability density of contaminant particle velocity for different system temperatures. Theoretical values are given by the Maxwell speed distribution for particles of a mass of 100 au at the respective temperature (Eq.~\ref{eq:maxwell}). We find exact agreement with the theory, verifying the correct operation of the coupling algorithm (Sec.~\ref{sec:method:sub:algorithm}). 
\label{fig:maxwell}}
\end{figure}

To evaluate the coupling approach described in section~\ref{sec:method:sub:algorithm} and ensure sound thermostatistic behaviour of the contaminant particles, the particle velocity statistics per collision event for a range of temperatures are measured. Parameters kept fixed are the ambient pressure $p=3$~Pa, a contaminant mass of $m_{\zeta}=100$~au and an atmosphere particle mass of $m_{\alpha}=2$~au. Figure~\ref{fig:maxwell} documents the exact agreement between the simulation velocity field and the theoretical solution

\begin{equation}
\label{eq:maxwell}
f(u) = 4\pi u^{2} \left( \frac{m}{2\pi k_{\text{B}}T} \right)^{3/2} \exp\left\{-\frac{m u^{2}}{2 k_{\text{B}}T}\right\}.
\end{equation}
 This result suggests correct thermalization of the \emph{pseudo-particle's} velocity components as well as functionality of the collision algorithm, serving in addition to the coupling as a thermostat to the contaminants.

\subsection{The second law of Fick - Diffusion in a binary mixture with large density contrast}
\label{sec:valFick}
As detailed above, a main focus of application of the proposed model is the efficient simulation of diffusive transport in low pressure environments. The validity of the event driven algorithm, parameterised by the mean free path $\lambda$ is tested by comparison of the measured resulting diffusivity. The second law of Fick
\begin{equation}\label{eq:diffPDE}
\frac{\partial}{\partial t} n_{\zeta}(x) = D_{\alpha\zeta} \nabla^{2} n_{\zeta}
\end{equation}
defines the contaminant-atmosphere diffusivity $D_{\alpha\zeta}$ in term of the second spatial and first time derivative of a concentration or density field. A solution in one dimension gives the particle density $n_{\zeta}(x,t)$ at a locus $x$ different from the initial position $x_{0}$ and time $t$ in terms of the absolute particle number $N_{\zeta}$ and $D_{\alpha\zeta}$ as
\begin{equation}\label{eq:fick2}
n_{\zeta}(x,t) = \frac{N_{\zeta}}{\sqrt{4\pi D_{\alpha\zeta} t}} \exp \left\{ -\frac{(x-x_{0})^{2}}{4 D_{\alpha\zeta} t} \right\}.
\end{equation}

\begin{figure}
\center
\includegraphics[width=0.5\textwidth]{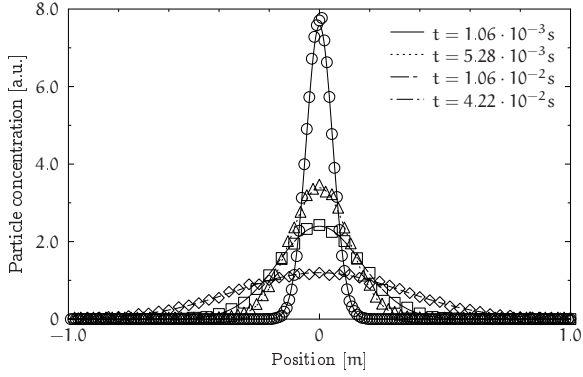}
\caption{Particle density distribution at different times. All particles have been initialised in the point of origin. We find quantitative agreement with the theory (Fick's second law, Eq.~\ref{eq:fick2})  over a wide range of parameters. Here depicted is an example configuration; a system comprised of a Hydrogen atmosphere at $p=3$~Pa and $T=295$~K containing contaminants of mass $m_{\zeta}=100$~au. The diffusivity is calculated according to Eqs.~\ref{eq:Diffusivity},~\ref{eq:massFactor} to $D_{\alpha\zeta}\approx1.31\cdot10^{-2}m^{2}/s$.\label{fig:diffusion}}
\end{figure}
Figure~\ref{fig:diffusion} illustrates the dynamics of the particle density field of a system comprised of a Hydrogen atmosphere at $p=3$~Pa and $T=295$~K containing contaminants of mass $m_{\zeta}=100$~au. For the presented results, 1 million particles are initially placed at the origin. Using a mass corrected diffusivity
\begin{equation}\label{eq:Diffusivity}
  D_{\alpha\zeta}=\mathcal{C}_{m}\frac{3}{8\sigma_{\alpha\zeta}^{2}n_{\alpha}} \sqrt{\frac{k_{B}T}{2\pi m_{\alpha\zeta}^{\star}}},
\end{equation}
with a mass correction factor
\begin{equation}\label{eq:massFactor}
\mathcal{C}_{m}=\frac{2m_{\alpha}+4m_{\zeta}}{3\left(m_{\alpha}+m_{\zeta}\right)},
\end{equation}
excellent quantitative agreement of simulation and theoretical predictions is found. The correction factor given by Eq.~\ref{eq:massFactor} has allowed quantitative predictions for the diffusivity over a wide range of atmosphere/contaminant mass contrasts. It can be motivated by the non-trivial expectation values of momentum transfer at large mass contrasts. Even for simplified models of hard spheres, the asymmetry in momentum carried by species of different mass, travelling at different respective thermal speeds, requires higher order mass corrections. For a more in-depth discussion and example calculations see the book of Chapman and Cowling~\cite{chapman_mathematical_1970}. The form of mass correction reported here is again depending on the simplifying assumptions limiting interaction to atmosphere-contaminant collisions. For a mass contrast of one $\mathcal{C}_{m}$ reduces to unity.

\subsection{The transport equation - The particle boundary condition}
\label{sec:valTransport}

\begin{figure}
\center
\includegraphics[width=0.5\textwidth]{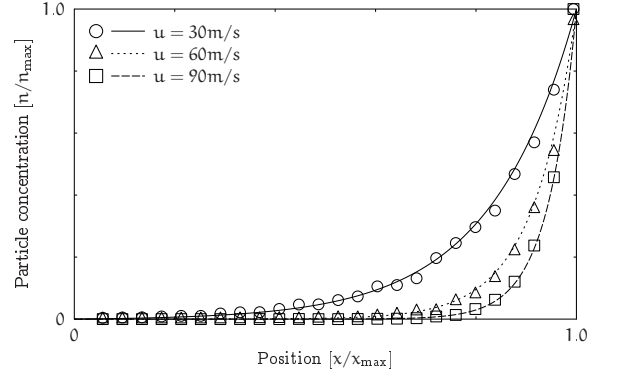}
\caption{Equilibrium particle density distribution in systems combining constant flow in positive x-direction and a diffusivity of $D_{\alpha\zeta}\approx1.31\cdot10^{-2}m^{2}/s$. At $x/x_{\textrm max}=1$ the system is delimited by a boundary acting on the contaminants (see section~\ref{sec:method:sub:algorithm}). The curves are given by the solution Eq.~\ref{eq:soltransport} to the transport equation assuming an infinite particle reservoir at $x>1$ and an open boundary at $x<0$. The simulation particle densities are re-normalised to the density measured 5 lattice sites in front of the wall. Outside of the boundary layer we find the differential equation~\ref{eq:transport} excellently approximated. The fluctuating densities due to thermalised reflection at the wall do however not justify the infinite reservoir assumption of the theory. This results in strong variation of the obtained result.
\label{fig:advdiff}}
\end{figure}

In this final evaluation of the model, the stationary state of a system described by the transport equation
\begin{equation}
\label{eq:transport}
\frac{\partial}{\partial t} n_{\zeta}(x) = -D_{\alpha\zeta} \nabla^{2} n_{\zeta} + u_{\alpha}(x) \nabla n_{\zeta} = 0
\end{equation}
is considered. Assuming the boundary conditions 
\begin{equation}
 n_{\zeta}(0) = 0\ ;\qquad n_{\zeta}(x/x_{\textrm max}=1) = 1,
\end{equation}
the particle number density is normalised by a maximum value assumed at $x=1$. This situation is describing a system with a constant flow velocity in positive x-direction and an infinite particle reservoir at $x>1$ as well as an open boundary at $x<0$. The infinite particle reservoir is here modelled by a boundary visible for the contaminant particles only. As before for the example system parameterisation a Hydrogen atmosphere at $p=3$~Pa and $T=295$~K containing contaminants of mass $m_{\zeta}=100$~au is selected. This corresponds to a diffusivity of $D_{\alpha\zeta}\approx1.31\cdot10^{-2}m^{2}/s$. Figure~\ref{fig:advdiff} depicts the normalised particle density over the normalised x-coordinate. Solid lines represent solutions to equation~\ref{eq:transport}, given by
\begin{equation}
\label{eq:soltransport}
n_{\zeta}(x) = \left(\exp\left\{ x \frac{u}{D_{\alpha\zeta}}\right\} - 1\right) \left(\exp\left\{ \frac{u}{D_{\alpha\zeta}}\right\} - 1\right)^{-1}.
\end{equation}
The simulation data is rescaled by the particle density 5 lattice sites in front of the wall. This allows to compare the data to theory applicable to an infinite reservoir. Measurements in closer vicinity to the wall suffer for the particle counts considered here from large density fluctuations introduced by the thermal wall boundary conditions.

\section{Simulation of contaminant suppression by bezels}\label{sec:contaminantTransport:contaminantSuppression}

The model is applied to a setup where contaminant suppression by low speed flows through different openings at low pressure is measured. The simulations have been performed in a two-step process. First, a solution to the flow of the background gas was obtained by means of an adjusted LBM. Contaminants were subsequently simulated coupled to the quasi-static solution.  In this part the LB parameterisation outlined in the preceding section~\ref{sec:contaminantTransport:parametrisation} is employed.

\subsection{System setup}

\begin{figure}
\begin{center}
  \includegraphics[width=0.5\textwidth]{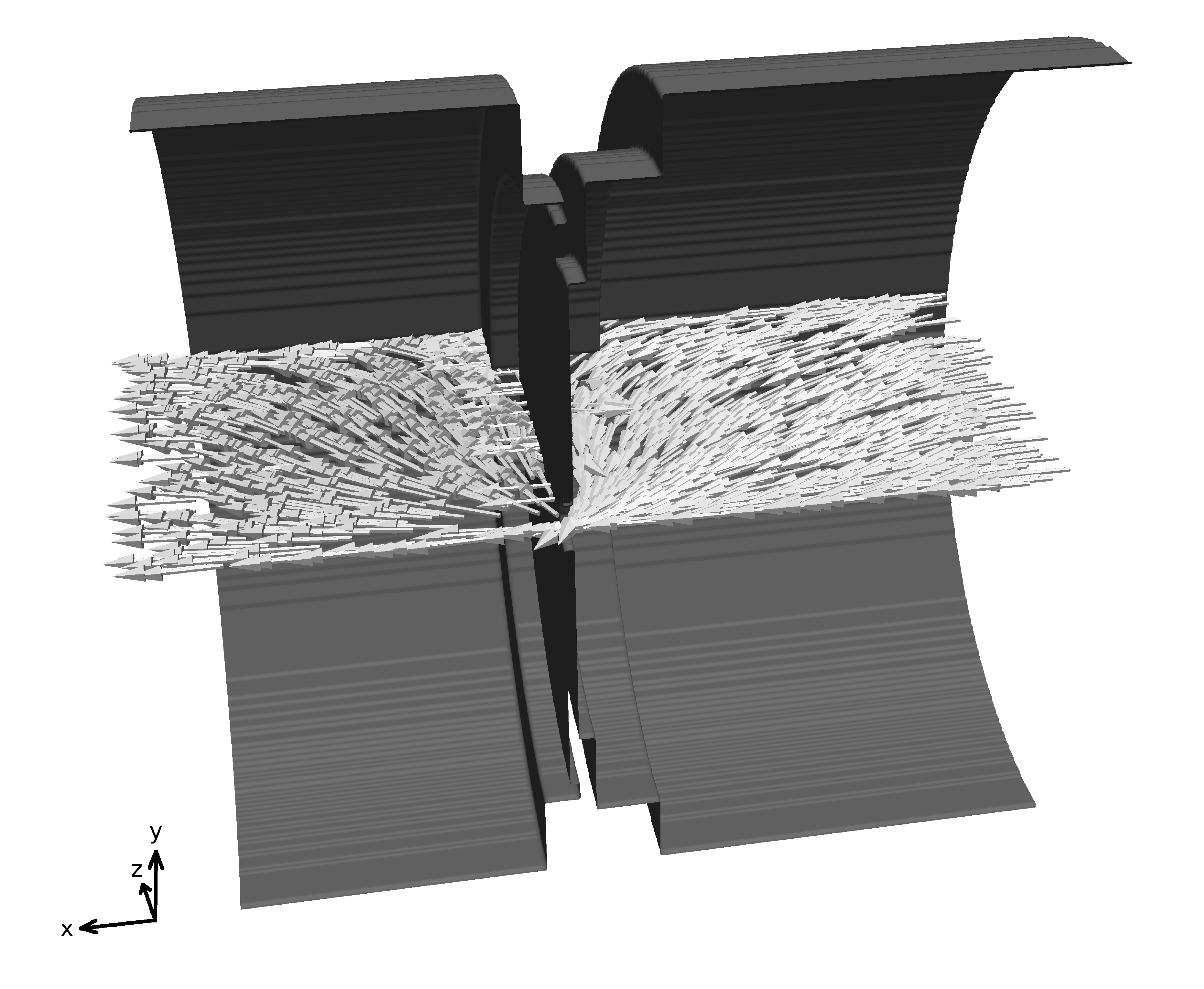}
\end{center}
\caption{Illustration of the flow path through the system. The velocities are not to scale. In the centre opening, the velocity is up to three orders of magnitude larger than in the remainder of the system.
\label{fig:bezelVelocityIllustration}}
\end{figure}

Figure~\ref{fig:bezelVelocityIllustration} shows a ray-traced image of the simulation geometry used. The cylindrical shape of the chamber is reflecting an experimental setup. The system is resolved by $256\times 256\times 256~{\textrm{l.u.}}^{3}$, where in the centre-plate a round opening with a diameter of $14~{\textrm{mm}}$ is placed.

The system geometry has been chosen to mimic a typical experimental measurement setup. Here, openings are introduced into the centre of a system with adjustable absolute pressure as well as relative pressure gradient. Flow rates are controlled by the pressure gradient and contaminants are injected on one side. Partial pressure is measured in both chambers.

Single component fluid flow is parameterised by the choice of the mean lattice density $\latticeDensity = 0.1~\latticeUnits$ and a relaxation time of $\relaxationTime = 5.3457~\latticeUnits$ At the given discretisation the viscosity wall-function~(\ref{eq:local-relaxation-time}) is found to have a non-zero value over 5 lattice sites.
Full refractive boundary conditions are employed in the immediate vicinity of the bezel opening only. This has been established to enhance the stability of the LB in regimes of higher pressure gradients. This is justified as, while introducing an error in the exact form of the flow field throughout the system, flow rates in the central opening are not affected by this measure. 

The flow is driven establishing a flow rate by a Neumann condition~\cite{zou_pressure_1997} on the influx at $x=0$ against a fixed pressure Dirichlet condition~\cite{zou_pressure_1997} at $x=n_{x}=256$. A snapshot illustration of the resulting flow field is depicted in figure~\ref{fig:bezelVelocityIllustration}. It is noteworthy that the flow rate in the region of the opening is up to three orders of magnitude higher than in the rest of the system, where vectors have been scaled equally to illustrate the flow path rather than the scales.

\begin{figure}
\includegraphics[width=0.5\textwidth]{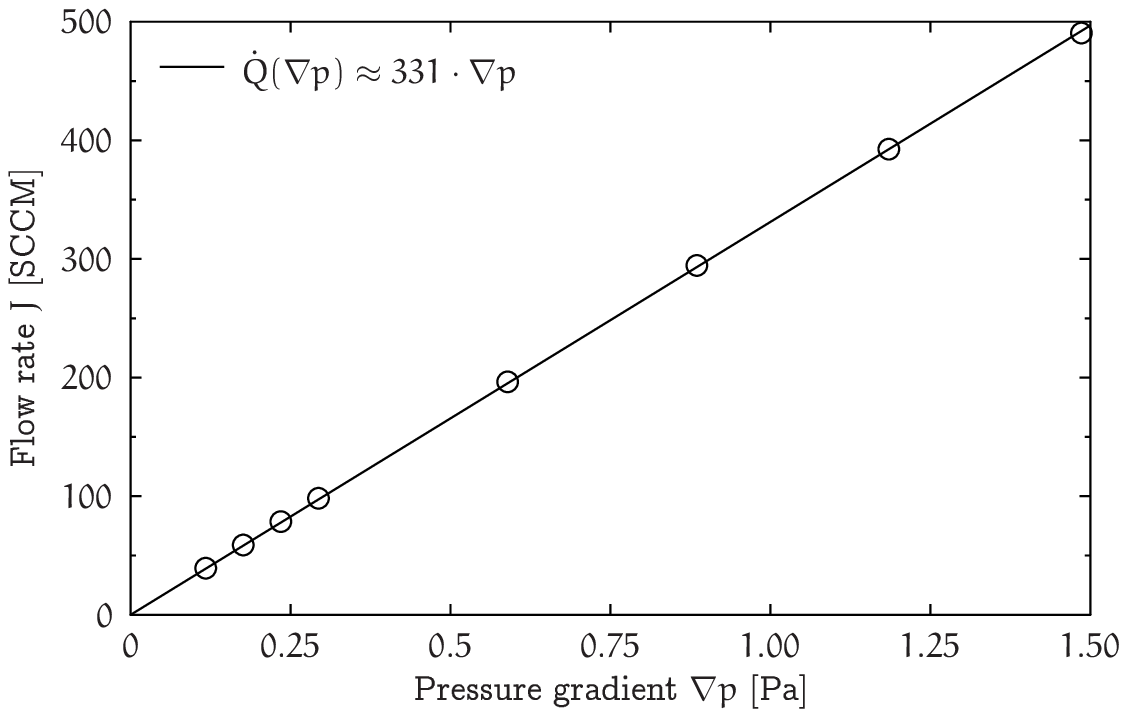}
\caption{Gas flow rate as a function of the pressure gradient. The LB simulation has been parameterised to the kinematic viscosity $\shearViscosity = 3.581 \frac{{\text {m}^2}}{\text s}$  and speed of sound $\speedOfSound = 1280 \frac{\text m}{\text s}$ of Hydrogen gas at a mean pressure of $\pressure = 4~{\text Pa}$ and temperature $\temperature = 295~{\text K}$. A clear linear dependence of the flow rate $\flowRate$ on the pressure gradient is found even at very high flow speeds violating the low Mach number assumption ($\mach \approx 0.27$).
\label{fig:bezelFlowrates}}
\end{figure}

In figure~\ref{fig:bezelFlowrates} atmosphere gas flow rates are plotted over the applied pressure gradients. The LB simulation has been parameterised to the kinematic viscosity $\shearViscosity = 3.581 \frac{{\text {m}^2}}{\text s}$  and speed of sound $\speedOfSound = 1280~\frac{\text m}{\text s}$ of Hydrogen gas at a mean pressure of $\pressure = 4~{\text Pa}$ and temperature $\temperature = 295~{\text K}$~\cite{national_institute_of_standards_and_technology_thermophysical_2011}. The mean free path of hydrogen in this setup is calculated from tabulated data of the collisional cross section and mass to be $\meanFreePath_{\alpha} \approx 2.70 \cdot 10^{-3} m$, implying a Knudsen number of $\knudsen \approx 0.2$ using the opening diameter as limiting length scale.  

Linear dependence of the flow rate $\flowRate$ as measured in the volume of the opening on the pressure gradient is found. Even though in the simulations exceeding a pressure gradient of $1~{\text Pa}$, the low Mach number limitation is clearly violated (at the highest value the Mach number in the peak flow is $\mach \approx 0.27$), the functional dependence holds surprisingly well. This is due to the fact that this violation only occurs for a few lattice sites. The LB simulations are run for 300,000 time steps after which the change in the field per time step is in the order of $10^{-7}$ and below.

\begin{figure}[!htbp]
\begin{center}
\includegraphics[width=0.5\textwidth]{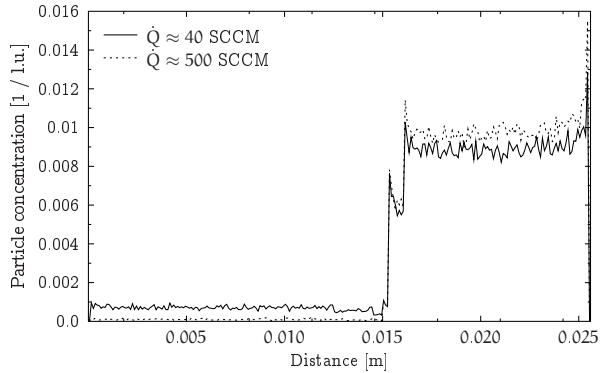}
\end{center}
\caption{Snapshot of simulation data at the lowest and highest flow rates, respectively 100,000 collision events into the simulation where a total of 10,000,000 collision events was simulated. The data is a histogram of the amount of particles present in the respective lattice layer volume. The total number of particles was 100,000. The suppression coefficient is calculated from fitting a constant function to these data in the ranges $x \in \left\{ 0.0050 .. 0.0100 \right\}$~m and $x \in \left\{ 0.0175 .. 0.0225 \right\}$~m. The step deviations in the central region reflect that the data is not normalised for the variation in local volume in the geometry. In particular, the opening in the region $x \in \left\{ 0.015 .. 0.0152 \right\}$~m is clearly visible as minimum. For the calculation of the suppression coefficient by Eq.~(\ref{eq:suppressionCoefficient}) particle numbers left and right of the opening are simply summed.
\label{fig:bezelSuppressionIllustration}}
\end{figure}

\subsection{Contaminant suppression}

The contaminant particles are initialised in the upstream chamber in the range of $x=0.0153$~m to $x=0.0256$~m. Each particle receives random spatial coordinates outside of the obstacle volume, as well as Maxwell-Boltzmann distributed random velocity components for a temperature of $\temperature = 295~{\text K}$ and particle mass of $100{\text au}$. The effective cross section is estimated from tabular values for Hydrogen and Heptane (as an example of a heavy organic molecule of a weight of 100 a.u.) to be~\cite{karniadakis_microflows_2000}
\begin{equation}
\sigma_{\alpha\zeta} \approx \frac{1}{2} \left( 2.91 \cdot 10^{-10} + 6.66 \cdot 10^{-10}\right) m = 4.785 \cdot 10^{-10} m,
\end{equation}
suggesting with equation (\ref{eqLambda}) a mean free path of the contaminant of approximately $\meanFreePath_{\zeta} \approx 1.98 \cdot 10^{-4} m$. Using the bezel opening diameter as typical scale, the obtained Knudsen number is $\knudsen \approx 0.014$. 

The contaminant dynamics are evaluated from concentration histograms of the system in the Cartesian coordinates. Figure~\ref{fig:bezelSuppressionIllustration} illustrates these data in the main flow direction for the minimum and maximum flow rates employed ($\flowRate \approx 40$~SCCM and $\flowRate \approx 500$~SCCM, respectively) after 100,000 of 10,000,000 collision events. The difference in the system dynamics is already at this early stage distinctly visible. The concentrations $\concentration^{+}$ and $\concentration^{-}$ used to determine the suppression coefficient as 
\begin{equation}\label{eq:suppressionCoefficient}
\suppressionCoefficient = \frac{\concentration^{+}}{\concentration^{-}}
\end{equation}
are measured by fitting a constant in regions of undisturbed concentrations in the ranges $x \in \left\{ 0.0050 .. 0.0100 \right\}$~m for the downstream and $x \in \left\{ 0.0175 .. 0.0225 \right\}$~m for the upstream chamber. The deviations in the concentrations in the central region reflect that the data is not normalised for the variation in local volume in the geometry. In particular, the opening in the region $x \in \left\{ 0.0150 .. 0.0152 \right\}$~m is clearly visible as minimum.

The suppression coefficients obtained for the various flow rates are presented in figure~\ref{fig:bezelSuppressionIllustration}. In agreement with findings involving an earlier model, an exponential dependence of the suppression on the flow rate is found~\cite{sinz_simulation_2008}. 

This test case illustrates the applicability of our hybrid model to real world systems with industrial relevance. Quantitative comparison with experimental data is a natural next step.
 
\begin{figure}
\begin{center}
\includegraphics[width=0.5\textwidth]{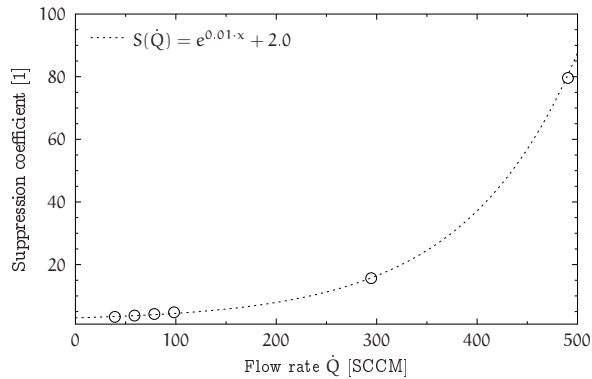}
\end{center}
\caption{Measured suppression coefficients over varying flow rate through the opening. The suppression behaviour of the system can be approximated by an exponential relation to the flow rate observed in the system.
\label{fig:bezelSuppression}}
\end{figure}

\section{Conclusion and Outlook}

Motivated by industrial applications requiring contaminant suppression in low vacuum environments, a hybrid simulation model has been refined and validated. A Monte Carlo algorithm to introduce a diffusivity model to a LBM setup has been developed and improved in thermostatistic context.

The problem considered comprises an ideal gas which is flushing a system pressurised in the single digit Pascal range, where medium sized organic contaminants are present in the system.
The sparsity of the contaminant molecules allows complete neglect of their impact on the gas flow. As a result it is necessary to couple the solution of the gas flow into the simulation of the contaminants only.

This important simplification allows efficient calculation of the diffusive transport (The contaminant simulations are executed as a single process and finish in a matter of hours). By only requiring a quasi static solution to the flow field, the Monte Carlo portion of the model is furthermore independent of the choice of flow solver used to obtain said solutions.

With the intermediate goal of simulating and predicting flow on scales relevant to engineering, the LBM has been selected due to its adaptability to complex geometries. While the method is not readily applicable to intermediate Knudsen flow regimes, phenomenological corrections can be introduced to recover the flow behaviour observed in rarefied systems with sufficient accuracy. 

The Monte Carlo model is distinguished by combination of two central ideas. First, the coupling of a local advective velocity by shift of the expectation values of the Gaussian thermal velocity distributions constituting the local Maxwellian. Second, the parameterisation of a collision event-driven algorithm determining time scales by the mean free path and equilibrium velocities.  

The resulting algorithm has been validated to recover correct thermal velocity distributions as well as superposed advective velocities. Since the coupling algorithm implicitly acts as a thermostat, the Maxwell-Boltzmann distribution of particle velocities is assured at all times.

Using collision properties of hard spheres with an additional mass contrast correction, quantitative prediction of the diffusivity of the contaminants for a given thermodynamical state and particle mass has been successful. The mass correction term found here may be of practical relevance to more than this application as it illustrates and quantifies the relevance of mass contrasts in momentum transfer and resulting diffusivity.

A more complex test case combines an effective advection-diffusion, or transport model with a test of the boundary conditions. It has to be refrained from developing a quantitative solution to the particle distribution in the boundary layer. Quantitative agreement with the approximation of an infinite fixed concentration source could however be obtained when a density integrated over the boundary layer was used.

In a final step the developed combined method has been applied to suppression of contaminant diffusive transport by gas flows through an opening in the low intermediate Knudsen regime. Qualitative evaluation of gas flow rate and resulting suppression has yielded promising results. As expected there exists a well defined linear dependence of the flow rate on the applied pressure gradient. Furthermore the resulting suppression is an exponential function of the flow rate.

Further development and application of the model can be envisioned in both experimental as well as theoretical context. True to the original motivation of the work a natural next step is the comparison with experimental results. To this end, this work provides a parameterisation framework both of the LBM and the Monte Carlo algorithm where input can be provided in SI units. While some limitations such as the low Mach number limit and the principal continuity assumption apply, the method is able to cover a wide range of technically relevant scales.

Theoretical extensions of the model can be made by means of refining the algorithm to increase its overall accuracy. A very obvious flaw of the current model living on the LBM lattice is the very limited resolution of boundary shapes. Also the coupling to models capable of providing exact solutions to the Knudsen layer flow to some problems could be very interesting. Also extension of the collision model to include soft potentials, etc. is of interest.

\acknowledgements{
Financial support is acknowledged from the TU Eindhoven and NWO/STW (Vidi grant 10787 of J. Harting). We thank the J\"{u}lich Supercomputing Centre and the High Performance Computing Center Stuttgart for the technical support and the CPU time which was allocated within large scale grants of the Gauss Center for Supercomputing.
}



%


\end{document}